\documentclass[aps,prc,preprint,groupedaddress]{revtex4}

\newcommand{\beq}{\begin{equation}}
\newcommand{\eeq}{\end{equation}}
\newcommand{\beqa}{\begin{eqnarray}}
\newcommand{\eeqa}{\end{eqnarray}}

\usepackage{epsfig}

\begin{document}

 \title{Sea-quark effects in the pion charge form factor}

\author{Q. B. Li}
\email[]{ligb@pcu.helsinki.fi}
\affiliation{Helsinki Institute of Physics
POB 64, 00014 University of Helsinki, Finland}

\author{D. O. Riska}
\email[]{riska@pcu.helsinki.fi}
\affiliation{Helsinki Institute of Physics, POB 64,
00014 University of Helsinki, Finland}

\thispagestyle{empty}

\date{\today}
\begin{abstract}
It is shown that the data on the pion charge form factor admit the
possibility for a substantial sea-quark components in the pion wave
function. If the charge form factor is calculated with instant form
kinematics in a constituent quark model that is extended to include
explicit $(q\bar q)^2$ components in the pion wave function, that
component will give the dominant contribution to the calculated
$\pi^+$ charge form factor at large values of momentum transfer. The
present experimental values $Q^2$ can be described well with $(q\bar
q)^2$ component admixtures of up to 50\%. The sensitivity of the
calculated $\pi^+$ charge form factor to whether one of the quarks
or one of the antiquarks is taken to be in the P-state is small.

\end{abstract}


\maketitle

\section{Introduction}

The small mass of the pion, in comparison to that of the vector
mesons, suggests that the pion is a collective state of $(q\bar
q)^n$ configurations, with many values of $n$. Therefore it is
something of a riddle that it is possible to describe the empirical
charge form factors of the charged pions satisfactorily with
phenomenological wave functions under the assumption that they are
pure quark-antiquark states \cite{coe}. It is then a natural
question to ask whether the empirical pion charge form factors can
exclude the presence of the expected multiquark configurations in
the pion.

The negative parity of the pion requires that in the simplest
sea-quark configuration, $(q\bar q)^2$, at most 3 of the
constituents can be in the ground state and that either one of the
quarks or one of the antiquarks is raised to the $P-$state (or a
higher odd$-L-$state). As this is energetically unfavorable it
suggests that the probability of that configuration may be small in
comparison to that of the $q\bar q$ component. The situation is
analogous to that in baryons, for which positive parity requires
that in a $qqqq\bar q$ admixture either one of the quarks or the
antiquark has to be in the $P-$state \cite{helm}.

Here the charge form factor of the (charged) pion is calculated in
an extension of the constituent quark model to include admixtures of
the simplest sea-quark configurations $(q\bar q)^2$ in instant form
kinematics. The calculation is made both for the case where one of
the quarks and where one of the antiquarks is in the $P-$state. It
is in both cases found that inclusion of the sea-quark configuration
allows a good description of the empirical form factor, even if it
represents as much as half of the wave function. The main point is
however, that as soon as there is a non vanishing probability for
the sea quark component in the wave function, that component will
lead to the dominant contribution to the charge form factor at large
values of momentum transfer in the case of instant form kinematics.
As a consequence the fact that it is possible to achieve a
quantitatively satisfactory fit to the empirical charge form factor
with wave function model for the conventional $q\bar q$ component
alone does not rule out the presence of significant sea-quark
components in the pion.

The method of calculating the charge form factor in the constituent
quark model with instant form kinematics developed here can readily
be extended to sea-quark configurations with larger numbers of
sea-quark $q\bar q$ components.

The configurations of the $(q\bar q)^2$ system that are possible in
the pion are described in section 2. Section 3 contains a
description of the pion wave function and the different form factor
contributions. The calculated results for the pion charge form
factor are given in section 4. Finally section 5 contains a
summarizing discussion.

\section{Light flavor $(q\bar q)^2$ configurations in the pion}

The lightest $(q\bar q)^2$ component in the pion meson
contains only the light flavor quarks $u$ and $d$, which form the
fundamental representation of the $SU(2)$ flavor symmetry.
The flavor-spin-color configurations of the $qq$ and $\bar q\bar q$
are listed in table \ref{qqqq} by their Young patterns.

\begin{table}
\caption{The flavor-spin-color configurations of
the $qq$ and $\bar q\bar q$ pairs. \label{qqqq}}
\begin{tabular}{l|c|c} \hline\hline
                 &   $qq$      &       $\bar q\bar q$  \\
\hline
$SU(2)_{flavor}$ & $[2]_F$,~$[11]_F$    &  $[2]_F$,~$[11]_F$     \\
$SU(2)_{spin}$   & $[2]_S$,~$[11]_S$    &  $[2]_S$,~$[11]_S $    \\
$SU(3)_{color}$  & $[2]_C$,~$[11]_C$    &  $[22]_C$,~$[211]_C $  \\ \hline\hline
\end{tabular}
\end{table}

The wave functions of the $qq$ and $\bar q \bar q$ pairs in the
flavor-spin-color-orbital space should be totally antisymmetrized,
respectively. In addition, the odd parity of the pion meson
requires that either a quark or a
antiquark in the $qq \bar q\bar q$ component is in P-state (or higher
odd$-L-$state.

In the case where one quark is in the P-state and both antiquarks are
in the S-state the flavor-spin-color wave functions of the
$qq$ pairs in the $(q\bar q)^2$
component are totally symmetric while that of the $\bar q\bar q$ pairs are
totally antisymmetric. This leads to four possible color singlet configurations
of the $(q\bar q)^2$ component  with
$J^P=0^-$:
\begin{eqnarray}
a.& \{[2]_F[2]_S[2]_C\}_{qq}\{[2]_F[11]_S[22]_C\}_{\bar q\bar q} \nonumber\\
b.& \{[2]_F[2]_S[2]_C\}_{qq}\{[11]_F[2]_S[22]_C\}_{\bar q\bar q} \nonumber\\
c.& \{[2]_F[11]_S[11]_C\}_{qq}\{[2]_F[2]_S[211]_C\}_{\bar q\bar q} \nonumber\\
d.& \{[11]_F[2]_S[11]_C\}_{qq}\{[2]_F[2]_S[211]_C\}_{\bar q\bar q}\,.
\label{abcd1}
\end{eqnarray}

It is natural to assume that the $(q\bar q)^2$ configuration
with the lowest energy shall have the
largest probability in the pion
besides that of the conventional $q\bar q$ component.
The splitting of the energy of the $(q\bar q)^2$ components (\ref{abcd1}) is determined by
the hyperfine interaction between the quarks and the antiquarks.
This interaction will here be taken to have the schematic form:
\begin{eqnarray}
H_I=-X\,\sum_{i\neq j}\vec\sigma_i\cdot \vec\sigma_j
\vec\lambda^c_i\cdot\vec\lambda^c_j\, .
\label{hi}
\end{eqnarray}
Here $\vec\sigma$ are the Pauli matrices, $\vec\lambda^c$ are the
Gell-Mann matrices in color space and $X$ is a positive constant
with the dimension of energy. This interaction has the same color
and spin dependence as the color magnetic hyperfine interaction,
which arises from single gluon exchange. The contributions to the
energies of the 4 configurations (\ref{abcd1}) that arise from the
schematic hyperfine interaction (\ref{hi}) can be determined by the
re-coupling method described in \cite{close} and \cite{jaffe} and
are listed in table \ref{expec1}.

\begin{table}
\caption{The expectation values
$-{1\over X}\langle \alpha\vert H_I\vert\alpha\rangle$ of the $(q\bar q)^2$
configurations in eq. (\ref{abcd1}).
\label{expec1}}
\begin{ruledtabular}
\begin{tabular}{c|cccc}
 $\alpha$  &  a &  b &  c  & d \\ \hline
 $-{1\over X}\langle \alpha\vert H_I\vert\alpha\rangle$ &
   -$8\over 3$  & 16 & $16\over 3 $ & 0  \\
\end{tabular}
\end{ruledtabular}
\end{table}

In the case where one antiquark is in the $P-$state, and both the
quarks are in the $S-$state the wave functions of the
$qq$ pairs in flavor-spin-color space have to be totally antisymmetric,
while those of the $\bar q\bar q$ pairs shall be totally symmetric.
The possible $(q\bar q)^2$ configurations in the pion are in
this case the following:
\begin{eqnarray}
a^\prime.& \{[2]_F[2]_S[11]_C\}_{qq}\{[2]_F[11]_S[211]_C\}_{\bar q\bar q} \nonumber\\
b^\prime.& \{[2]_F[2]_S[11]_C\}_{qq}\{[11]_F[2]_S[211]_C\}_{\bar q\bar q} \nonumber\\
c^\prime.& \{[2]_F[11]_S[2]_C\}_{qq}\{[2]_F[2]_S[22]_C\}_{\bar q\bar q} \nonumber\\
d^\prime.& \{[11]_F[2]_S[2]_C\}_{qq}\{[2]_F[2]_S[22]_C\}_{\bar q\bar q}\,.
\label{abcd2}
\end{eqnarray}
The expectation values of the hyperfine interaction between quarks
in the $(q\bar q)^2$ configurations in (\ref{abcd2}) are listed
in table \ref{expec2}. The results in the tables \ref{expec1}
and \ref{expec2} show that the hyperfine interaction between quarks leads
to the same energy levels of the $(q\bar q)^2$
configurations, in which the antiquark is in the S-state and
P-state. This is a consequence of the
fact that the hyperfine interaction
is independent of the angular momentum of the constituent quarks.
In the case, where the antiquark is in the P-state the $(q\bar q)^2$
configuration $d^\prime$ has the lowest energy, which is equal to
that of the lowest energy
$(q\bar q)^2$ configuration $b$, in which the antiquark is in the S-state.
These two configurations are thus likely to constitute the most
probable $(q\bar q)^2$
components and therefore to be most significant in the structure of the
pions.
The roles of such configurations in the pion charge form factor
will be considered in the following section.

\begin{table}
\caption{The expectation values
$-{1\over X}\langle \alpha^\prime\vert H_I\vert\alpha^\prime\rangle$
of the $(q\bar q)^2$
configurations in eq. (\ref{abcd2}).
\label{expec2}}
\begin{ruledtabular}
\begin{tabular}{c|cccc}
 $\alpha^\prime$  &  $a^\prime$ &  $b^\prime$ &  $c^\prime$  & $d^\prime$ \\ \hline
 $-{1\over X}\langle \alpha^\prime\vert H_I\vert\alpha^\prime\rangle$ &
   $16\over 3$  & 0 & -$8\over 3 $ & 16  \\
\end{tabular}
\end{ruledtabular}
\end{table}

\section{$\pi^+$ charge form factor}

\subsection{The form factor contribution from the $q\bar q$ component}

The present empirical results for the
charge form factor of the  $\pi^+$ may be well described as a
$q\bar q$ system in a Poincar\'e covariant constituent quark model
with the following orbital wave function model \cite{coe}:
\begin{eqnarray}
\phi({\vec k}_1,{\vec k}_2)={\cal N}_{2q}
{1\over (1+{\sum_{i=1}^2{{\vec k}_i}^2\over 2b^2})^{a}}\,.
\label{owave2q}
\end{eqnarray}
Here ${\cal N}_{2q}$ is a normalization factor, $a$ and $b$ are
parameters and $k_i$, $i$=1, 2, are the quark momenta in the rest
frame of $\pi$ meson ($\sum_{i=1}^2{\vec k}_i$=0). This wave
function model will be adopted here for the $q\bar q$ component of
the pion wave function.

In the impulse approximation the contribution of the $q\bar q$ component to the
pion charge
form factor is obtained as
the matrix element of the electric current
density operator between the initial and final states
in the Breit frames as:
\begin{eqnarray}
F_\pi^{(q\bar q)}(Q^2)=\int d^3{\vec p}_2 \,
\sqrt{J_2 J_2^\prime} \, S_e({\vec p}_1,{\vec p}_1\,^\prime)\,
\phi({\vec k}_1,{\vec k}_2)
\phi({\vec k^\prime}_1,{\vec k^\prime}_2)\, .
\label{2qform}
\end{eqnarray}
The initial and final momenta of the constituents in their
respective Breit frames are denoted
$\vec p_i$ and $\vec p_i\,^{'}$ respectively ($\vec p_1\,^{'}=\vec p_1+\vec Q$ and
$\vec p_2=\vec p_2\,^{'}$).

If the momentum transfer $\vec Q$ is taken to define the $z-$axis, the
relations in instant form kinematics between the momenta in the
Breit frames and the rest frames are:
\begin{eqnarray}
&&\vec  p_{i\perp} = \vec k_{i\perp}
 = \vec k_{i\perp}\,^{'} = \vec{ p}_{i \perp}\,^{'}\, ,
\nonumber \\
&&
 p_{i\parallel} = v_{0} k_{i\parallel}+v_{\parallel}\,
\omega_i \, ,  \nonumber \\
&&
 p_{i\parallel}^{'} = v_{0}^{'} k_{i\parallel}^{'}
+v_{\parallel}^{'}\, \omega_i^{'} \, ,  \nonumber \\
&&E_i= v_{\parallel}k_{i\parallel}+v_{0}\omega_i\, ,
\nonumber\\
&&E_i^{'}=v_{\parallel}^{'}k_{i\parallel}^{'}
+v_{0}^{'}\omega_i^{'}\, .
\label{boosts}
\end{eqnarray}
Here the energy components are defined as
\begin{eqnarray}
&&\omega_i=\sqrt{\vec k_i\,^2 + m^2}\,,\quad
\omega_i^{'}=\sqrt{\vec k_i\,^{'2} + m^2}\, ,\nonumber\\
&&E_i=\sqrt{\vec p_i\,^2 + m^2}\, ,\quad
E_i^{'}=\sqrt{\vec p_i\,^{'2} + m^2}\, .
\label{energies}
\end{eqnarray}
In these relations $m$ denotes the constituent mass
and $v=\{v_{0},\vec 0_\perp,v_{\parallel}\}$ and
$v'=\{v_{0}^{'},\vec 0_\perp,v_{\parallel}^{'}\}$ the
constituent boost velocities in the initial and final
states. These satisfy the constraint $v^2=v^{'2}=-1$.

In instant form kinematics the boost velocities may be
defined as \cite{Bruno}:
\begin{eqnarray}
&&v_{\parallel}=-{Q\over 2\sum_{i=1}^n \omega_i}\, ,
\nonumber\\
&&v_{\parallel}^{'}=\,\,\,\,{Q\over 2\sum_{i=1}^n \omega_i^{'}}\, .
\label{Vboosts}
\end{eqnarray}
Here $n$ represents the number of constituents.

The Jacobian that is induced by the transformation between the rest
frame and the Breit frame of the meson is in the case of the $q\bar
q$ component obtained as \cite{ours}:
\begin{eqnarray}
J_2&=&{\omega_2\over E_2}(1-v_{\parallel}
{k_{1\parallel}\over E_1})\, .
\label{Jac2}
\end{eqnarray}
The expression for the corresponding final state Jacobian
$J_2^{'}$ (\ref{2qform})is obtained by replacement of the arguments by the
corresponding primed coordinates.

The electric
current density operator in eq. (\ref{2qform}) is
\begin{eqnarray}
S_e(\vec p, \vec p^\prime)=\sqrt{1+{Q^2\over 4 M_\pi^2}}
\sqrt{{(E'+m)(E+m)\over 4 E' E}}
\bigg\{1+{\vec p\,'\cdot \vec p\over (E'+m)(E+m)}
\bigg\}\, .
\label{elasticJ}
\end{eqnarray}

The expressions for the boost velocities of the constituents
(\ref{Vboosts}) reveal, that their magnitudes fall with increasing
number of constituents $n$, if the constituent mass is constant.
Given that form factors fall with increasing momentum transfer $Q^2$
it follows that, at sufficiently large $Q^2$, the wave function
component that contains the largest number of constituents will give
the largest contribution to the form factor. This feature is
explicit in instant form kinematics. It has a natural physical
interpretation in that the form factor  describes the probability
that the system stays bound upon absorption of the momentum transfer
$Q$. The relative probability for this to happen is smaller if few
constituents absorb the momentum transfer than if many can share it
so that the fractional momentum transfer per constituent is smaller.

\subsection{The form factor contribution from the $(q\bar q)^2$ component}

In the case where both of the  antiquarks in the $(q\bar q)^2$
component are in the $S-$state, the
wave function of the $(q\bar q)^2$ component, which has the
symmetry configuration $b$ in eq. (\ref{abcd1}), and which is expected
to have the lowest energy, may be expressed as:
\begin{eqnarray}
|\pi^+\rangle_S = -{1\over \sqrt{2}}\,uu(\bar d \bar u - \bar u \bar d)\,
\{[{\bf 1}_S\otimes {\bf 1}_S^\prime]_1\otimes {\bf 1}_X\}_{0^-}\,
\{{\bf 6}_C\otimes {{\bf \bar 6}}_C\}_{1_C}\,
\Phi({\vec k}_1,{\vec k}_2,{\vec k}_3, {\vec k}_4)\, .
\label{ppwave1}
\end{eqnarray}
Here $\vec k_i$, $i$=1...4, are the momenta of the constituent
quarks in the rest frame of the pion ($\sum_{i=1}^4{\vec k}_i=0$).
The spin triplet combinations of the $qq$ and $\bar q\bar q$ pairs
are denoted ${\bf 1}_S$ and ${\bf 1}_S^\prime$, respectively. These
combine  with the P-state $qq$ pairs to the total quantum numbers of
pion $J^P=0^-$. The spherical harmonic for the $qq$ pairs in the
$(q\bar q)^2$ component is defined as
\begin{eqnarray}
{\bf 1}_X,_m={\xi_1}_m\,, \qquad \vec\xi_1=
{1\over\sqrt{2}}({\vec k}_1-{\vec k}_2)\, ,
\end{eqnarray}
where ${\xi_1}_m$ ($ m=-1, 0, 1$) are the spherical components of $\vec\xi_1$.
In (\ref{ppwave1}) we have denoted the Young pattern
representations of the color
states of $qq$ and $\bar q\bar q$ pairs in table \ref{qqqq} with their
corresponding dimensions (${\bf 6}$).

The orbital wave function of the $(q\bar q)^2$ component is taken
to have the form:
\begin{eqnarray}
\Phi({\vec k}_1,{\vec k}_2,{\vec k}_3, {\vec k}_4)=
{\cal N}_{4q}
{1\over (1+{\sum_{i=1}^4{{\vec k}_i}^2\over 2B^2})^{A+1}}\, ,
\label{owave4q}
\end{eqnarray}
where ${\cal N}_{4q}$ is a normalization factor.

Then the contribution to the $\pi^+$ charge form factor from
the $qq\bar q \bar q$ component in the case where one of the
antiquarks is in the S-state then may be written as:
\begin{eqnarray}
F_{\pi^+}^S(Q^2)={4\over 3}A^S(Q^2)-{1\over 3}B^S(Q^2)\, .
\label{form}
\end{eqnarray}
Here the terms $A$ and $B$ are  defined as:
\begin{eqnarray}
A^S(Q^2)={1\over 3}\int d^3{\vec p}_2 d^3{\vec p}_3 d^3{\vec p}_4
\sqrt{J_4 (1) J_4^\prime(1)}  S_e({\vec p}_1,{\vec p}_1^\prime)
{\vec\xi}_1\cdot{\vec\xi}_1^\prime
\Phi({\vec k}_1,{\vec k}_2,{\vec k}_3, {\vec k}_4)
\Phi({\vec k^\prime}_1,{\vec k^\prime}_2,{\vec k^\prime}_3, {\vec k^\prime}_4) ,
\label{asq2} \\
B^S(Q^2)={1\over 3}\int d^3{\vec p}_1 d^3{\vec p}_2 d^3{\vec p}_3
\sqrt{J_4 (4) J_4^\prime (4)}  S_e({\vec p}_4,{\vec p}_4^\prime)
{\vec\xi}_1\cdot{\vec\xi}_1^\prime
\Phi({\vec k}_1,{\vec k}_2,{\vec k}_3, {\vec k}_4)
\Phi({\vec k^\prime}_1,{\vec k^\prime}_2,{\vec k^\prime}_3, {\vec k^\prime}_4) .
\label{bsq2}
\end{eqnarray}
Here $A^S (Q^2)$
represents the matrix element where a
photon couples to the 1-st quark in the $(q\bar q)^2$
component, while $B^S (Q^2)$
represents the matrix element where the
photon couples the antiquark (the fourth constituent).

The Jacobians for the transformations between the corresponding
Breit frames and the rest frames are:
\begin{eqnarray}
J_4(1)&=&{\omega_2\omega_3\omega_4
\over E_2 E_3 E_4}(1-v_{\parallel}
{k_{1\parallel}\over E_1})\, ,\\
J_4(4)&=&{\omega_1\omega_2\omega_3
\over E_1 E_2 E_3}(1-v_{\parallel}
{k_{4\parallel}\over E_4})\, .
\label{Jac4}
\end{eqnarray}
As in the case of eq. (\ref{2qform}), the primed variables in eqs.
(\ref{asq2}) and (\ref{bsq2}) represent the final states variables
that correspond to the initial state variables without primes. Here
we do not take into account the Wigner rotation of the spin axis
that is caused by the boosts, as its consequences are numerically
insignificant for momentum transfers below 10 GeV$^2$
\cite{Bruno,CoRi}. Comparison of the expressions for the Jacobians
(\ref{Jac2}) and (\ref{Jac4}) for the case of 2 and 4 constituents,
respectively, makes it clear how to generalize these expressions to
the case of $n$ constituents.

In the case where one antiquark is in the $P-$state and both quarks
are in the ground state, the wave function of the $(q\bar q)^2$
component with the lowest energy, which has the symmetry
configuration $d^\prime$ in eq. (\ref{abcd2}) may be obtained from
that in the case where one of the quarks are in the $P-$state
(\ref{ppwave1}) by the replacements:
\begin{equation}
u\leftrightarrow -\bar d \quad d\leftrightarrow \bar u\, .
\label{replace}
\end{equation}
The explicit expression is then
\begin{eqnarray}
|\pi^+\rangle_P = {1\over \sqrt{2}}(ud-du)\bar d \bar d \,
\{[{\bf 1}_S\otimes {\bf 1}_S^\prime]_1\otimes
{\bf 1}^\prime _x\}_{0^-}\,
\{{\bf 6}_C\otimes {\bar {\bf 6}}_C\}_{1_C}\,
\Phi({\vec k}_1,{\vec k}_2,{\vec k}_3, {\vec k}_4)\, .
\label{ppwave2}
\end{eqnarray}
Here the spherical harmonic for the $\bar q\bar q$ pair
is denoted as
\begin{eqnarray}
{{\bf 1}^\prime}_x,_m={\xi_3}_m\, ,\qquad \vec\xi_3=
{1\over\sqrt{2}}({\vec k}_3-{\vec k}_4)\, .
\end{eqnarray}

The explicit expression for the $\pi^+$ charge form factor in the
case where one antiquark is in the $P-$state are is:
\begin{eqnarray}
F_{\pi+}^P(Q^2)={1\over 3}A^P(Q^2)+{2\over 3}B^P(Q^2) \, .
\label{formp}
\end{eqnarray}
Here the orbital integrals are defined as:
\begin{eqnarray}
A^P(Q^2)={1\over 3}\int d^3{\vec p}_2 d^3{\vec p}_3 d^3{\vec p}_4
\sqrt{J_1(1) J_1^\prime (1)}  S_e({\vec p}_1,{\vec p}_1^\prime)
{\vec\xi}_3\cdot{\vec\xi}_3^\prime
\Phi({\vec k}_1,{\vec k}_2,{\vec k}_3, {\vec k}_4)
\Phi({\vec k^\prime}_1,{\vec k^\prime}_2,{\vec k^\prime}_3, {\vec k^\prime}_4) ,\\
B^P(Q^2)={1\over 3}\int d^3{\vec p}_1 d^3{\vec p}_2 d^3{\vec p}_3
\sqrt{J_4(4) J_4^\prime(4)} S_e({\vec p}_4,{\vec p}_4^\prime)
{\vec\xi}_3\cdot{\vec\xi}_3^\prime
\Phi({\vec k}_1,{\vec k}_2,{\vec k}_3, {\vec k}_4)
\Phi({\vec k^\prime}_1,{\vec k^\prime}_2,{\vec k^\prime}_3, {\vec k^\prime}_4) .
\label{abq2p}
\end{eqnarray}
The symmetrical form of the expressions (\ref{ppwave1}) and
(\ref{ppwave2}) has the consequence that
\begin{eqnarray}
A^P(Q^2)=B^S(Q^2)\, ,~~~~~~~~~~~B^P(Q^2)=A^S(Q^2)\, .
\label{equal}
\end{eqnarray}

Because of the symmetry structure of the spin-flavor-color state of
the $(q\bar q)^2$ components, there is no contribution of
off-diagonal $(q\bar q)^2\rightarrow q\bar q$ transition matrix
elements to the pion charge form factor. In the case of the nucleons
the contribution of such transition matrix elements to the form
factors are much larger than that of the corresponding diagonal
matrix elements\cite{ours}.

\begin{figure}[t]
\vspace{20pt}
\begin{center}
\mbox{\epsfig{file=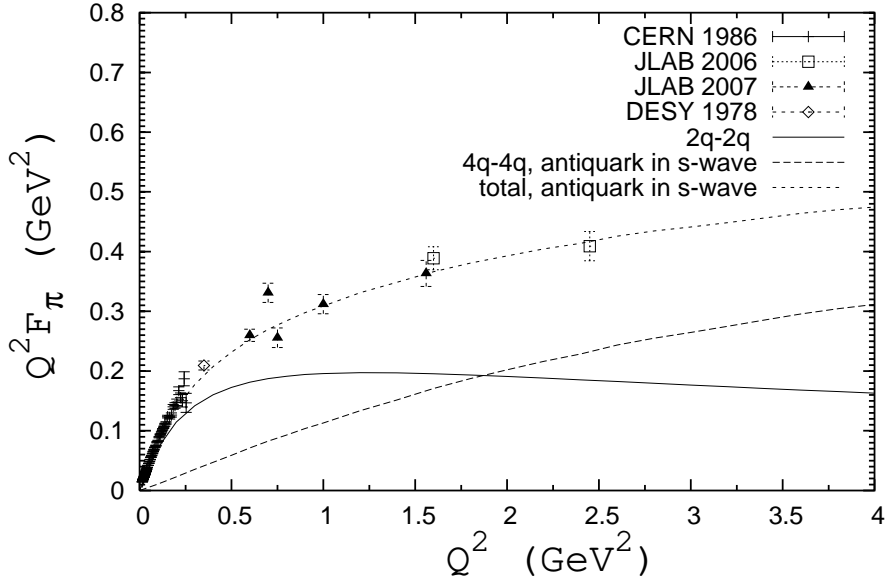, width=120mm}}
\caption{The $\pi^+$ charge form factor obtained with a 10\%
 $(q\bar q)^2$ component probability
with both antiquarks in the $S-$state. Solid line: The contribution
from the $q\bar q$ component; Dashed line: The contribution from the
$(q\bar q)^2$ component; Dotted line: the result combining the
contributions from the $q\bar q$ and the $(q\bar q)^2$ components.
The data sets CERN 1986, DESY 1978, JLAB 2006 and JLAB 2007 are
taken from refs. \cite{cern1986,desy1978,jlab2006,jlab2007},
respectively. The data point at $Q^2=1.60 $GeV$^2$ from ref.
\cite{jlab2007} has been shifted for better visibility.}
\label{pform}
\end{center}
\vspace{10pt}
\end{figure}

\section{results}

To investigate possible role of the $(q\bar q)^2$ component in the
form factor of the $\pi^+$ meson, the wave function parameters are
chosen so that the combined contribution of the $q\bar q$ and the
$(q\bar q)^2$ components yield a form factor that agrees with the
empirical one under the assumption of a probability for the latter
component of 10 \%. The corresponding wave function parameters $a,b$
(\ref{owave2q}) and $A,B$ (\ref{owave4q}) are listed in table
\ref{parameters}. The constituent quark mass was taken to be 120
MeV, which is close to the value required to describe the nucleon
form factors in instant form kinematics with a wave function of
corresponding form \cite{Bruno}. The wave function parameters for
the $q\bar q$ and the $(q\bar q)^2$ components were chosen such,
that the empirical mean square radius for the pion, $r_\pi^2 = 0.44$
fm$^2$, was recovered.

\begin{table}
\caption{The model parameters.\label{parameters}}
\begin{ruledtabular}
\begin{tabular}{lccccc}
$(q\bar q)^2$& $m_q$ (MeV)  & $b$ (MeV) & $B$ (MeV) &$a$ &$A$ \\
\hline

10\%&  120       &  190        &   100 & 2.3 & 1.8  \\
20\%&  120       &  190        &   110 & 2.3 & 2.0  \\
40\%&  120       &  190        &   139 & 2.3 & 2.21 \\
50\%&  120       &  190        &   143 & 2.3 & 2.25 \\
\end{tabular}
\end{ruledtabular}
\end{table}

The calculated result for the $\pi^+$ charge form factor is shown in
Fig. \ref{pform} for the case where both antiquarks are in the
$S-$state in the $(q\bar q)^2$ component. The result indicates that
above 1 GeV$^2$ that with these parameters the main form factor
contribution arises from the smaller $(q\bar q)^2$ component. That
this should be so is in fact quite natural, as in the case of
elastic form factors, the form factor falloff with momentum should
depend on $Q^2$ divided by the square of the number of involved
constituents. In this case the contribution of the $(q\bar q)^2$
component is very small (and in fact negative): -0.03 fm$^2$. The
sign of this contribution depends on the parameter values.

\begin{figure}[t]
\vspace{20pt}
\begin{center}
\mbox{\epsfig{file=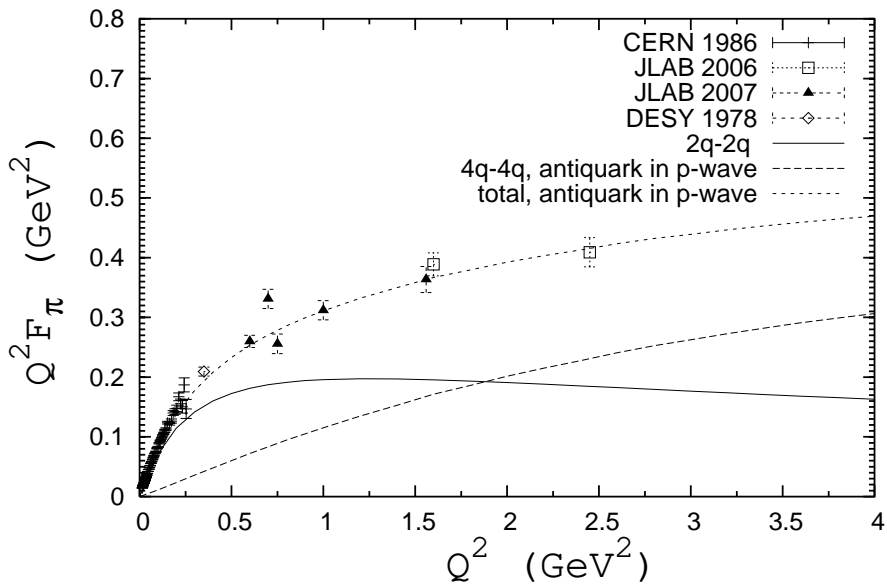, width=120mm}}
\caption{The $\pi^+$ charge form factor with a 10\% probability for
the $(q\bar q)^2$ component with one antiquark in the $P$-state. The
labeling of the curves is the same as that in Fig. \ref{pform}.}
\label{pformp}
\end{center}
\vspace{10pt}
\end{figure}

The corresponding calculated results for the $\pi^+$ charge form
factor for the case, where one of the antiquarks is in the $P-$state
are shown in Fig. \ref{pformp}. These results are in fact very
similar to those obtained in the former case, where both antiquarks
are in the $S-$state, the main difference being a slightly larger
magnitude for the contribution to the mean square radius from the
$(q\bar q)^2$ component in the present case (-0.04 fm$^2$).

\begin{figure}[t]
\vspace{20pt}
\begin{center}
\mbox{\epsfig{file=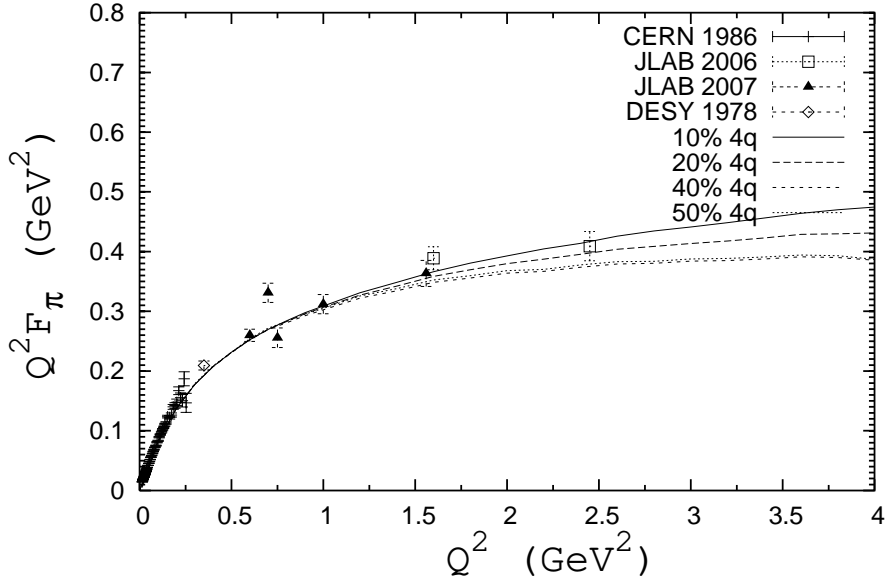, width=120mm}}
\caption{The $\pi^+$ charge form factor obtained with a 10\%-50\%
 $(q\bar q)^2$ component probabilities
with both antiquarks in the $S-$state. The data sets CERN 1986, DESY
1978, JLAB 2006 and JLAB 2007 are taken from refs.
\cite{cern1986,desy1978,jlab2006,jlab2007}, respectively. The data
point at $Q^2=1.60 $GeV$^2$ from ref. \cite{jlab2007} has been
shifted for better visibility.} \label{pformallN}
\end{center}
\vspace{10pt}
\end{figure}

In Fig. \ref{pformallN} the form factor is shown as obtained for
different values of the probability for the $(q\bar q)^2$ component.
These results were obtained by only slight variation of the 2
parameters in the wave function of the $(q\bar q)^2$ component
(\ref{owave4q}). These results show that the present empirical data
on the pion charge form factor can allow for a $(q\bar q)^2$
component probability of up to 50\%.

\begin{figure}[t]
\vspace{20pt}
\begin{center}
\mbox{\epsfig{file=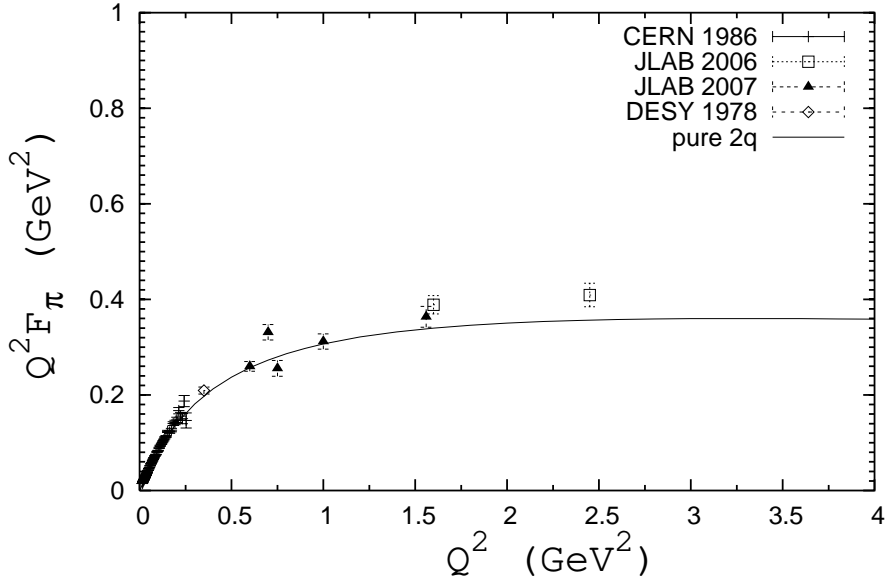, width=120mm}}
\caption{The $\pi^+$ charge form factor for the pure $q\bar q$
model. The labeling of the curves is the same as that in Fig.
\ref{pform}.} \label{pform2q}
\end{center}
\vspace{10pt}
\end{figure}

For comparison the form factor that is obtained for the pure $q\bar
q$ quark model for the pion is shown in Fig. \ref{pform2q}. These
results were obtained with the constituent quark mass value 80 MeV
and with the parameters $a=2.0$ and $b=198$ MeV in the $q\bar q$
wave function model (\ref{owave2q}). This shows that the pion charge
form factor may be described with such a simple model wave function
in instant form kinematics, a result that was noted for the case of
front form kinematics in ref.\cite{coe}.

\begin{figure}[t]
\vspace{20pt}
\begin{center}
\mbox{\epsfig{file=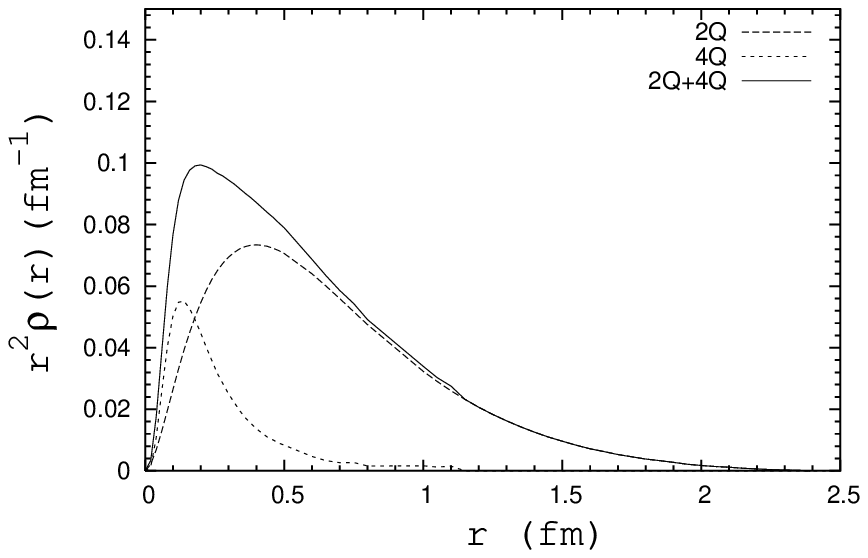, width=120mm}}
\caption{The "charge density" of the $\pi^+$ for a $(q\bar q)^2$
component with 20\% probability. The contributions from the $q\bar
q$ and $(q\bar q)^2$ components are denoted $2Q$ and $4Q$
respectively.} \label{rho20}
\end{center}
\vspace{10pt}
\end{figure}

\section{conclusions}

The results of this study shows that the present data for the charge
form factor of the charged pions may described as well with
inclusion of sea-quark configurations with probabilities up to least
50 \% in the covariant constituent quark model with instant form
kinematics as without such configurations.

When the sea-quark component is included in the form factor, that
component will give the dominant contribution to the form factor at
sufficiently large $Q^2$, which dominates over that from the $q\bar
q$ component, however small its probability. The $(q\bar q)^2$
component corresponds to structures that have shorter range than the
basic $q\bar q$ component. This is illustrated in Fig.\ref{rho20},
where the charge density contributions from the $q\bar q$ and the
$(q\bar q)^2$ components, along with their sum, is show for the
case, in which the probability of the latter component is 20\%. If
the probability of the $(q\bar q)^2$ component is taken to be
larger, the peak in the profile $r^2\rho(r)$ moves towards that of
the $q\bar q$ component.

The fact that the sea-quark contribution gives the largest
contribution to the form factor at large values of momentum transfer
is a consequence of the fact that the magnitude of the energy
denominator in the expressions for the boost velocities for the
constituents (\ref{Vboosts}) grows with the number of constituents.
This implies that the momentum transfer is shared by the largest
number of constituents, and in effect, to a smaller relative
momentum transfer per constituent. As the form factor is a
monotonically falling function of momentum transfer, the consequence
is that the largest contribution at large $Q^2$ is given by the
component with the largest number of constituents.

\end{document}